\documentclass[twoside]{ilcws07}
\usepackage[latin1]{inputenc}
\usepackage[dvips]{graphicx,epsfig,color}
\usepackage{wrapfig,rotating}
\usepackage{amssymb,amsmath,array,url}

\def\ti              {\tilde}
\def\nt              {\ti\chi^0}
\def\to              {\rightarrow}
\def\surn            {SU(5)_{\rm RN}}
\begin{document}
\title{
%%%%   Paper title goes here  %%%%%%%%%%%%%%
Distinguishing SUSY scenarios using $\tau$ polarisation and $\nt_1$ Dark Matter.} %% 
%***********************************************************************
% AUTHORS INFORMATION AREA
%***********************************************************************
\author{L. Calibbi$^1$, R. Godbole$^2$, Y. Mambrini$^3$ and  S. K. Vempati$^2$.
% Optional short acknowledgment: remove next line if non-needed
%\thanks{This is an optional funding source acknowledgment.}
% DO NOT MODIFY THE FOLLOWING '\vspace' ARGUMENT
\vspace{.3cm}\\
% Addresses and institutions (remove "1- " in case of a single institution)
1- Departament de F\'{\i}sica Te\`orica, Universitat de
Val\`encia-CSIC, E-46100, Burjassot, Spain.
\vspace{.1cm}\\
2- Centre for High Energy Physics, Indian Institute of Science, 
Bangalore, 560012, India.
\vspace{.1cm}\\
3- Laboratoire de Physique Theorique, Universite Paris Sud, 
F-91405 Orsay, France\\
}
%%***********************************************************************
% END OF AUTHORS INFORMATION AREA
%***********************************************************************

\maketitle

\begin{abstract}
We discuss first a method of measuring $\tau$ polarisation at the ILC
using the $1$--prong  hadronic decays of the $\tau$.
We then show in this contribution how a study of the $\tilde \tau$ sector
and particularly use of decay $\tau$ polarisation can offer a very good 
handle for distinguishing between mSUGRA and a SUSY-GUT scenario, both of 
which can give rise to appropriate Dark Matter.
\end{abstract}

\section{Introduction}
Supersymmetry (SUSY)~\cite{Drees:2004jm} at the TeV scale provides one of  
the most attractive solution 
to the problem of instability of the Higgs mass under radiative correction. In 
fact SUSY forms the template of the physics beyond the Standard Model 
(BSM physics) that one wishes to probe at the coming colliders like the LHC 
and the ILC~\cite{Weiglein:2004hn}. In the $R$--parity conserving version of 
the theory, SUSY also provides a natural dark matter candidate, the lightest 
neutralino $\nt_1$. However, a consistent TeV scale supersymmetry is possible
with quite  different theoretical realisations at the high scale. For example, 
mSUGRA~\cite{Drees:2004jm} and SUSY-GUTs with/without seesaw 
mechanism~\cite{Calibbi:2007bk} are two models embodying SUSY with quite 
different high scale physics, both of which in turn  provide a satisfactory 
explanation of the Dark Matter (DM) in the Universe. The high scale physics 
of course leaves its imprints on the properties of the sparticles at the 
electroweak (EW) scale. The issue of being able to distinguish between such 
different scenarios using collider experiments has been a matter of great 
interest to the community. ILC with the possibilities of the high precision 
measurements offers itself as a natural candidate for the job in hand.  
In this contribution we show how a study of $\ti\tau$ 
sector can offer a good possibility of distinguishing the above mentioned
specific scenarios in the $\ti \tau$ -- $\nt_1$ co-annihilation region.

\section{$\tau$ polarisation: measurement and use as a SUSY probe.}
Recall that the mass eigenstates $\ti \tau_i, i=1,2$ and $\nt_j, j=1,4$ 
are mixtures
of ${\ti \tau}_L, {\ti \tau}_R$ and gauginos, higgsinos respectively. The 
couplings of a sfermion with a gaugino does not involve a helicity flip 
whereas that with a higgsino does. As a result the net helicity of  the $\tau$
produced in the decay $\ti \tau_i \to \nt_j \tau$, can carry information
about $L$--$R$ mixing in the $\ti \tau$ sector as well as that in the $\nt_j$
sector~\cite{Nojiri:1994it}.  In collinear approximation for the $\tilde\tau$ 
decay; i.e.  $m_\tau \ll m_{\tilde \tau_1}$, the polarisation of the $\tau$ 
produced, for example, in $\ti \tau_1 \to \tau \nt_1$ is given by,

\begin{eqnarray}
\mathcal{P}_\tau &=& \frac{\left(a^R_{11}\right)^2 - \left(a^L_{11}\right)^2}{
\left(a^R_{11}\right)^2 + \left(a^L_{11}\right)^2}; \nonumber
\\[2mm]
a^R_{11} &=& - \frac{2g}{\sqrt{2}} N_{11} \tan\theta_W
\sin\theta_\tau - \frac{gm_\tau }{\sqrt{2} m_W \cos\beta} N_{13}
\cos\theta_\tau, \nonumber \\[2mm]
a^L_{11} &=& \frac{g}{\sqrt{2}} \left[N_{12} + N_{11}
\tan\theta_W\right] \cos\theta_\tau - \frac{gm_\tau}{\sqrt{2} m_W
\cos\beta} N_{13} \sin\theta_\tau,
\end{eqnarray}
where we have used the standard notation~\cite{Drees:2004jm} with
the matrix $N$ representing the 
diagonalising matrix of the neutralino mass matrix
with the notation
$
\tilde \chi_1 = N_{11} \tilde B + N_{12} \tilde W + N_{13} \tilde H_1 +
N_{14} \tilde H_2.
$
$\mathcal{P}_\tau$ depends on the  mixing in the slepton sector as well 
as that in the neutralino sector which are determined by the SUSY model 
parameters; thus giving a good handle of the measurement of SUSY parameters.
%If, for example, the LSP is a $\tilde B$ as in mSUGRA then $P_\tau \simeq = 1$ 
%over the entire parameter range, whereas if  the LSP  dominated by the 
%Higgsino component over most of the parameter space, as is the case with
%some nonuniversal SUGRA models one expects $P_\tau \simeq \cos^2 \theta_\tau 
%- \sin^2 \theta_\tau.$

Even more importantly, $\tau$ polarisation can be also measured well at the 
colliders.  The energy distribution of the $\pi$ produced in the decay,
$\tau \to \nu_\tau \pi$ as well as those in $\tau \to \rho \nu_\tau,
\tau \to a_1 \nu_\tau$ depends on the handedness of the $\tau$.
In fact the angular distribution of the decay meson depends on $\tau$  
polarisation and is different for longitudinal and transverse states of
the vector meson $v$.
The  transverse (longitudinal) vector mesons share the energy of parent
meson evenly (unevenly) among the decay pions. For the $\tau$ decay the 
only measurable momentum is $\tau$-jet momentum and its value relative to 
$p_\tau$ is  determined by the meson decay angle. Hence the energy 
distribution of decay pions can be used then to measure the $\tau$ 
polarisation~\cite{Hagiwara:1989fn,Roy:1991sf,Bullock:1992yt}. As a matter 
of fact a lot of nice analysis of $\tau$ polarisation and hence of the MSSM 
parameter determination at a Linear Collider, making use of the 
$\tau \to \rho/ a_1 \nu_\tau$ (multi-prong) mode 
exist~\cite{Nojiri:1996fp,Boos:2003vf}. 

In this note we first discuss a method to determine the $P_{\tau}$ using
$1$--prong $\pi$ final state~\cite{Godbole:2004mq}.  
If we consider the inclusive distributions 
of the $1$--prong $\pi$ final state and define 
$R = p_{\pi^\pm} /  p_{\tau-jet}$, one finds that  
for $\mathcal{P}_\tau = 1$ the distribution in $R$ 
is peaked at $R < 0.2$ and $R  > 0.8$, whereas for 
$\mathcal{P}_\tau =-1$ it is peaked
in the middle. The observable $R$ can be simply determined by measuring
the energies of the $\tau$ in the tracker and the calorimeter. Further, 
the fraction 
$$
f  = {\sigma (0.2 < R < 0.8) \over \sigma_{total}},
$$ can be shown to be very nicely correlated with the
$\tau$--polarisation~\cite{Godbole:2004mq} and hence can be used as 
its measure.
%Since the $1$--prong inclusive hadronic decay of the $\tau$ corresponds to 
%$80\%$ of the hadronic decay and $50\%$ of total width, using this
%final state clearly increases the statistics of the sample used for 
%the determination of the $\tau$ polarisation. 
Note that full 
reconstruction of the $a_1$ and $\rho$ as needed in the mutli-prong
analysis is also not needed. 
 
\begin{figure}[t]
\centerline{
\includegraphics*[scale=0.27]{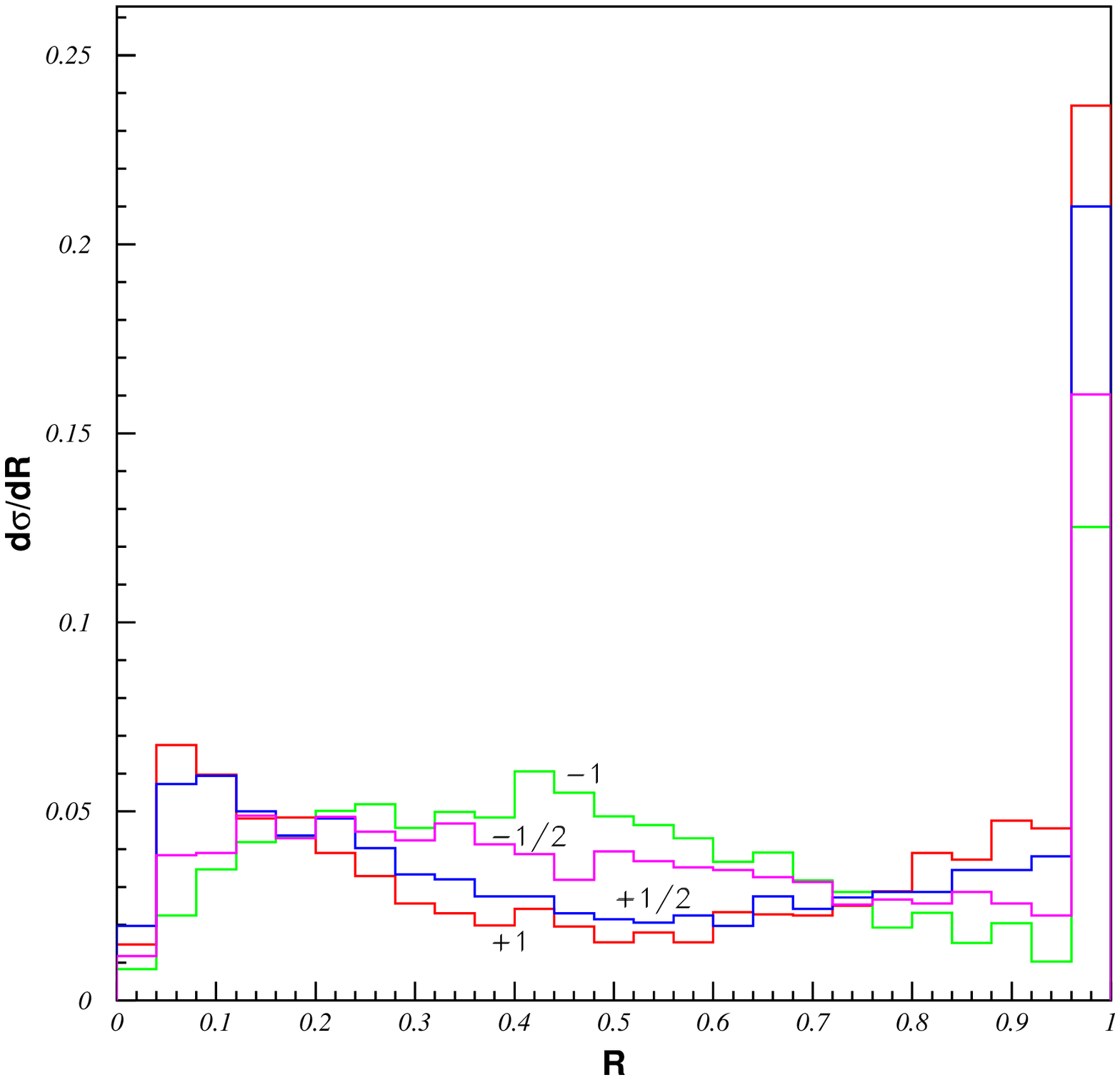}
\includegraphics*[height=5cm,width=5cm]{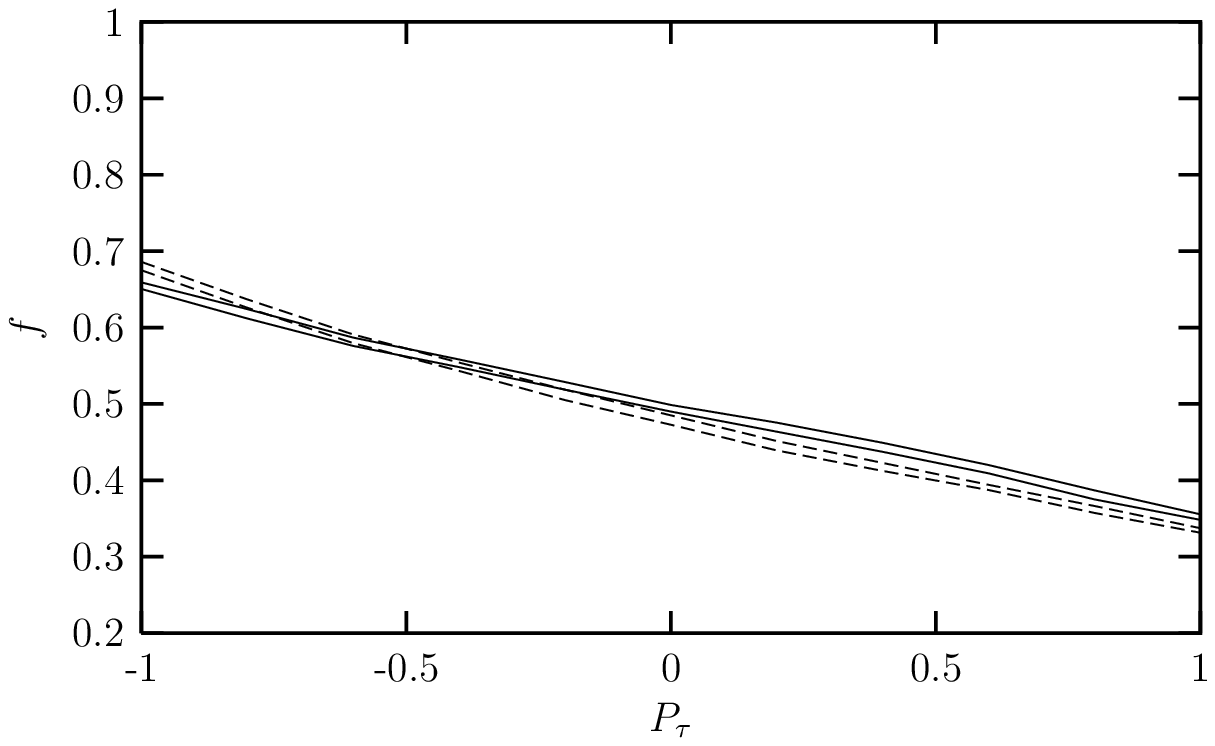}
}
\caption{Left panel shows $R$ distribution for $\tau$ produced in $\ti\tau_1
\to \tau \nt_1$ for different values of $\mathcal{P}_\tau$ and right panel shows
$f$, defined in text, as a measure of the polarisation. For details 
see~\protect\cite{Godbole:2004mq} 
\label{Fig:taupol} }
\end{figure}

The left panel in Fig.\ref{Fig:taupol} (taken from ~\cite{Godbole:2004mq}
shows distribution in $R$ for different values of  polarisations  
$\mathcal{P}_\tau$ as indicated on the figure for specific choice of 
$\sqrt{s}$, 
$\ti\tau_1,\nt_1$ masses and kinematical cuts on $\tau$ mentioned therein.
The right panel shows $f$ as a function of $\mathcal{P}_\tau$. 
Uncertainty due to the different parameterisations of the $a_1$ 
and non-resonant contributions to the $\pi$, give rise to the slight spread 
of the lines.  One can see from the Figure that 
$\Delta \mathcal{P}_\tau = \pm 0.03 (\pm 0.05)$ for 
$\mathcal{P}_\tau = -1 (+1)$. 
Even if an additional error were to come from the experimental measurement of 
$f$, still  a measurement of $\mathcal{P}_\tau$ with less than  $10 \%$ error,
i.e.  $\Delta \mathcal{P}_\tau < 0.1$ is sure to  be  possible. There 
is some dependence of the slope on the kinematics of the $\tau$, but it is 
clear from the  
%For
%$p^T_{\tau-jet}>$25 GeV cut shown by the solid lines, $f$ changes from 0.65
%to 0.35 for $P_\tau = -1$ to 0.35 at $P_\tau = +1$.  For
%$p^T_{\tau-jet} > 50$~GeV  shown by the dashed line  decrease is steeper.
figure that the use of inclusive 1-prong channel, 
is a robust method of determining $\tau$ polarisation. If the aim is only to
determine $\tau$ polarisation, then the $1$--prong method 
has the advantage of higher statistics and smaller systematic errors,
compared to the exclusive channel.

\section{SUSY-GUTs, mSUGRA and $\tau$ polarisation}
\begin{figure}[t]
\centerline{
\includegraphics*[width=0.33\textwidth]{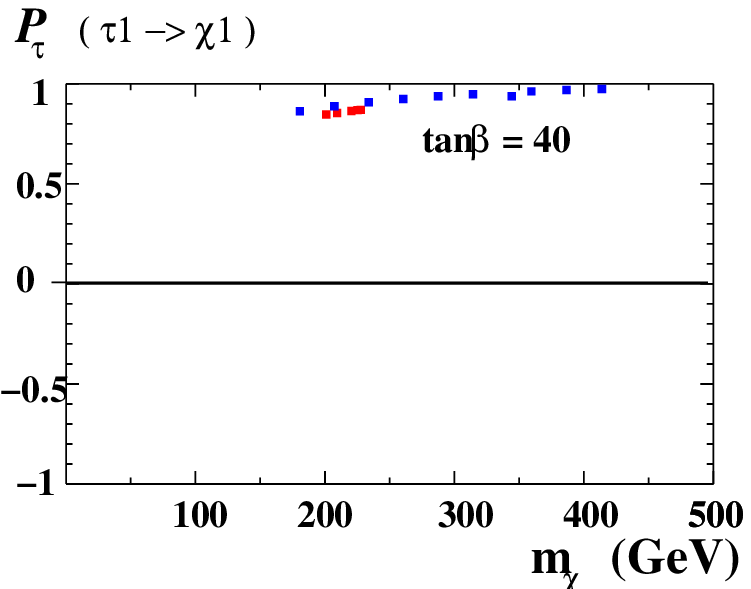}
\includegraphics*[width=0.45\textwidth]{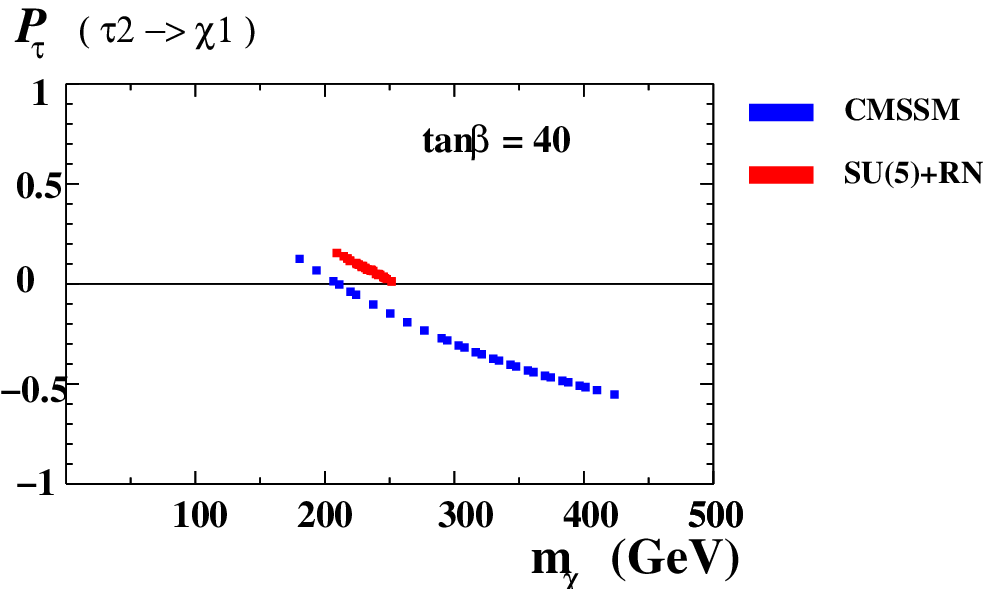}
}
\caption{Left panel shows $\mathcal{P}_\tau$ for mSUGRA and $\surn$ model in
$\tilde{\tau}_1 \to \nt_1$ decay as a function of the $\nt_1$
mass. The blue(dark) points are for mSUGRA whereas the red(grey) points are
for $\surn$ model. The right panel shows the same 
for $\tilde{\tau}_2 \to \nt_1$.
\label{ptautau1}
 }
\end{figure}

Let us now  see how the properties of the $\ti \tau$ sector and particularly
the $\tau$ poalrisation can be used to distinguish  between
various SUSY models. In the present case, we will choose mSUGRA
model and a SUSY $SU(5)$ with seesaw mechanism ($\surn$)~\cite{Calibbi:2007bk}. 
Requiring neutralino
DM relic density to be consistent with the recent WMAP measurements
significantly reduces the degeneracies present in the parameter space between 
these two models.  In fact, the effect is quite dramatic; in contrast to 
mSUGRA, the SUSY-GUT model has only two ``allowed" regions: (a) the 
\textit{stau} coannihilation channel, whose shape is quite different to the
corresponding one in mSUGRA\footnote{And further predicts an upper bound
on the $\nt_1$ mass for a given tan$\beta$.}; 
(b) the A-pole funnel region which does exist for large value of $\tan\beta$
whereas a focus point region is not present at least up to 5 TeV in the 
SUSY masses~\cite{Calibbi:2007bk}.
From the above it's clear that probing the $\tilde{\tau}$-neutralino 
sector could give a handle in distinguishing both the models as long
as SUSY spectrum is determined by the coannihilation region, where 
the masses of $\ti\tau_1$ and $\nt_1$ are very close. In fact,
in our analysis~\cite{ourpaper}, we find that the two models can be clearly 
distinguished from measuring $\mathcal{P}_\tau$ in the
decays of $\tilde{\tau}_2 \to \tau \nt_1$ (Fig.~\ref{ptautau1} right panel). 
Here for most of the parameter
space, the $\mathcal{P}_\tau$ has different signs. In the small overlap
region, $|\Delta \mathcal{P}_\tau|\gtrsim 0.2$, which make them 
distinguishable at the ILC. In the decay, 
$\tilde{\tau}_1 \to \tau \nt_1$, 
the tau polarisation cannot be really used to distinguish between both the 
models as we see from the left panel of Fig.~\ref{ptautau1}. 
In our analysis, we have assumed that $\tilde{\tau}_1$ and $\tilde{\tau}_2$ can
be distinguished from the kinematics (as $\tilde{\tau}_1$ is closer to mass
of $\nt_1$ in the coannihilation region). 

The behaviour of $\mathcal{P}_\tau$ from $\tilde{\tau}_2 \to \tau \nt_1$ 
in the two frameworks can be understood as follows. 
In the approximation of $\nt_1 \approx \tilde{B}$, very well
satisfied in the $\tilde{\tau}$ coannihilation region, $\mathcal{P}_\tau$
just depends on the $L$--$R$ mixing for $\tilde{\tau}$ and  
is simply related to the parameters entering the $\tilde{\tau}$ mass 
matrix~\cite{Nojiri:1994it}; given by:  
$$
\mathcal{P}_\tau = \frac{4 m^4_{\rm LR}-
(m^2_{\rm LL} -m^2_{\tilde{\tau}_1} )^2}
{4 m^4_{\rm LR}+ (m^2_{\rm LL} -m^2_{\tilde{\tau}_1} )^2}.
$$
Here  $m^2_{\rm LL}$ is the soft SUSY-breaking mass of 
$\tilde{\tau}_{\rm L}$, $m^2_{\tilde{\tau}_1}$ the lightest $\tilde{\tau}$
mass eigenvalue and $m^2_{\rm LR} \simeq -m_\tau \mu \tan\beta$ 
the $L$--$R$ mixing term.
From this expression, we can see that
the condition for a positive polarization reads:
\begin{equation}
\mathcal{P}_\tau > 0~ \Leftrightarrow ~2  \left|m^2_{\rm LR}\right|
> \left|  m^2_{\rm LL} -m^2_{\tilde{\tau}_1} \right| 
\label{polpos}
\end{equation}
In mSUGRA, such condition can be satisfied for a small region of the paramater 
space. The factor 2 on the l.h.s. of Eq.~\ref{polpos} plays a crucial role. 
In $\surn$, the mixing term $\left|m^2_{\rm LR}\right|$ is enhanced 
as an effect of RH neutrinos and GUT~\cite{Calibbi:2007bk}, and this
tends to make $\mathcal{P}_\tau$ larger. 
Moreover there is an upper bound on the $\nt_1$ mass in the coannihilation 
region: these two effects conspire to keep  
$\mathcal{P}_\tau$ always positive (right panel of Fig.~\ref{ptautau1}). 

Thus in this contribution we show how, using $\tau$ polarisation and $\nt_1$ 
DM constraints, we can go a long way in distinguishing various SUSY
models at the ILC. 
\\

\textbf{Acknowledgements} 

\noindent 
The work of L.C. is supported by the foundation ``Angelo Della Riccia'', 
and he also acknowledges the financial support of the spanish MEC and FEDER 
under grant FPA2005-01678.
The work of Y.M. is sponsored by the PAI programm PICASSO under 
contract PAI--10825VF and he would like to thank the European Network of
Theoretical Astroparticle Physics ILIAS/N6 under contract number
RII3-CT-2004-506222 and the French ANR project PHYS@COL\&COS for
financial support.

% ****************************************************************************
% BIBLIOGRAPHY AREA
% ****************************************************************************

\begin{footnotesize}
% IF YOU DO NOT USE BIBTEX, USE THE FOLLOWING SAMPLE SCHEME FOR THE REFERENCES
% ----------------------------------------------------------------------------

\end{footnotesize}
\end{document}